\documentclass{scrartcl}
\usepackage[numbers]{natbib}
\usepackage{url}
\usepackage{xcolor}
\usepackage{multicol}
\usepackage[a4paper, total={6.5in, 9.7in}]{geometry}
\usepackage{graphicx} 

\title{New Frontiers in Fighting Misinformation}
\subtitle{Exploring new strategies for tackling misinformation more effectively}

\author{Harith Alani and Grégoire Burel}
\date{Knowledge Media Institute, The Open University, UK}
\begin{document}
\maketitle

\begin{multicols}{2}

Despite extensive research and development of tools and technologies for misinformation tracking and detection, we often find ourselves largely on the losing side of the battle against misinformation. In an era where misinformation poses a substantial threat to public discourse, trust in information sources, and societal and political stability, it is imperative that we regularly revisit and reorient our work strategies. While we have made significant strides in understanding how and why misinformation spreads, we must now broaden our focus and explore how technology can help realise new approaches to address this complex challenge more efficiently. 

\paragraph{Amplifying Corrective Information\\}
Extensive research has focused on capturing and tracking how misinformation disseminates across social media platforms. Some studies have even extended the scope to incorporate data from reliable sources, allowing for comparisons between the spread of ``true'' and ``false'' information \cite{Vosoughi}. However, there has been significantly less focus on comprehending the concurrent dissemination of misinformation and its corrective counterpart, such as fact-checks \cite{gburel_2021}. For example, when tracking Covid-19 misinformation and their corresponding fact-checks on Twitter/X, we found the ratio of the appearance of misinformation in posts to be 3 times more than the appearance of their fact-checks \cite{gburel_2021}. Unfortunately, although social platforms are rightly pressured to remove misinforming content, there is no equivalent strong push to actively promote and advantage reliable and corrective information.  

Data analysis methods, such as spread variance and causal analysis, help to gain deeper insights into the dynamics between misinformation and their corrective information, and the influence they exert on each other. Such data-driven insights can lead to 
constructing fine-tuned strategies for disseminating corrective information in ways that maximise impact on misinformation spread.

\paragraph{Promoting Credible Accounts\\}

Evaluating the overall credibility of social media accounts can be a challenging task and requires extended observation of their activities over time. 
Although certain platforms have implemented measures to flag or remove misinformation content and occasionally block accounts for policy violations, there is a notable lack of effort and strategy dedicated to assessing the overall credibility of users and promoting or demoting them accordingly. In fact, the conventional algorithms employed for recommending users to follow could unintentionally boost the visibility of popular accounts that disseminate misinformation \cite{Tommasel}.

The pressing need for new strategies, policies, and developments in assessing, promoting, and demoting social media accounts based on their overall credibility is evident, as current approaches remain inadequate and constrained by data access limitations. While there are no universally accepted indicators of credibility, indices such as NewsGuardtech.com and Iffy.news, along with the output of legitimate fact-checking organizations worldwide, provide valuable resources for new algorithms to evaluate the credibility of social media users. 
Sharing debunked misinformation or content published by known unreliable sources reduces a user’s credibility while sharing reliable content from credible sources enhances it. Such indicators can be automatically detected and used to regularly compute and monitor the overall credibility of a given account \cite{guay2023}. However, the challenge in developing such technology is gaining appropriate access to users' data, and convincing social media platforms to utilise such techniques in their account recommendation algorithms.


\paragraph{Pre-empting Misinformation\\}

Pre-bunking misinformation is not a new concept. As a pre-bunking strategy, exposure to flawed argumentation techniques or weakened doses of misinformation was proved effective in increasing people's resistance to misinformation \cite{cook}. 
The fundamental premise of pre-bunking is that misinformation shares common traits and follows recurring templates. 
For example, misinformation during natural disasters often takes a similar shape, such as depicting sharks swimming in flooded areas or falsely portraying fires in regions that are actually safe. Claims that consuming a particular type of fruit can miraculously cure a serious disease are common health misinformation. In the context of war-related misinformation, it is now common to encounter claims that genuine footage is falsified or to assert that irrelevant or fake footage is from the actual war zone. 

To enhance our pre-emptive actions, we could produce technologies for \textit{predicting} the emergence of misinformation in evolving events and contexts. 
Knowledge Graphs with semantically structured and interlinked misinformation data, such as ClaimsKG (\url{github.com/claimskg}) and MisinfoKG (\url{github.com/CIMPLE-project/}) could be holding the key for identifying complex misinformation templates and semantic and temporal patterns. Coupling such knowledge with machine learning could enable early prediction of misinformation in recurring scenarios and thus countermeasures can be prepared and launched in advance.  

By monitoring patterns and predicting and recognising common themes, researchers can develop more advanced algorithms and strategies to identify and counteract misinformation in its infancy.




\paragraph{Optimal Timing for Corrections\\}
To maximize the impact of corrective information, we need a deeper understanding of when is the optimal time to intervene. Identifying the tipping point for correcting a claim and when it is better to ignore or delay it to prevent undue attention is important for our misinformation correction strategies. One lab experiment revealed disparities in the effects of presenting corrective messages before, during, or after exposure to misinformation \cite{brashier}. 
Such research emphasises the need for strategic timing in addressing misinformation, as corrections can be ineffective or even worsen the situation by backfiring or inadvertently amplifying false claims.

Research showed that the production of fact-checks has a notable impact on misinformation circulation and that this impact varies significantly depending on the specific topic of the misinformation \cite{gburel_2021}. Time-series analysis and predictive modelling can support the development of methods and strategies to accurately anticipate the tipping points for specific instances or topics of misinformation and thus enable more precise and effective timing for the release and boosting of corresponding fact-checks.


\paragraph{Real-World Experimentation\\}
 
Correcting misinformation is a multifaceted challenge influenced by a range of psychological, social, and technical factors. Almost all existing research that seeks to identify effective approaches is grounded in methodologies such as crowdsourcing, questionnaires, and lab-based simulations \cite{Gwia}. Translating lab-based findings into real-world settings, where individuals willingly and freely share misinformation, is an area that is largely unexplored. Therefore, the applicability and real-world effectiveness of lab-tested corrective interventions may be limited or uncertain. However, conducting experiments to test misinformation correction methods in real-world, natural settings is undeniably more difficult compared to controlled lab studies. Complexity arises from the multitude of variables that can influence the outcomes, many of which are challenging to control or measure in these dynamic, uncontrolled environments. It would also be important to address the resistance to misinformation corrections that will be encountered in real-world settings. 

This gap in our knowledge underscores the need for research that bridges the divide between controlled experiments and the dynamic landscape of social media. The replication of lab experiments in real-world settings is heavily dependent on obtaining suitable access to data from social media platforms and the cooperation and consent of these platforms to facilitate such experiments.



\paragraph{Audiences Do Matter\\}

Research on correcting misinformation has primarily centred around the individual, where various approaches are tested to alter a person's belief in a false claim or improve their behaviour towards misinformation. 
However, in social media, people and posts are likely to have a considerable audience, such as the followers of an account or a particular hashtag. 
The audience may not only consume the initial misinformation post but is likely to also witness any corrective actions taken, including the posting of corrective messages. They may further observe reactions to the misinformation and its corrections, which could come from the sender of the misleading post or other individuals. Moreover, this audience might actively partake in the discourse, engaging through actions such as posting, liking, or sharing.

This highlights the necessity of placing a stronger focus on investigating the audience's take from, and reactions to, interventions aimed at countering \textit{third-party} misinformation. Furthermore, it calls for an expansion of our misinformation-correction performance metrics to include audience responses in our assessments. In the context of social media, an intervention that may not successfully correct an individual's belief in a false claim but effectively deters bystanders from further sharing or correcting their own beliefs can still be deemed a positive outcome.

\paragraph{Long-Term Exposure to Misinformation\\}
It is natural to assume that exposure to misinformation is detrimental in some ways. 
Most relevant studies involved questioning participants before and after exposing them to misinformation to gauge the influence of misinformation on their choices and reasoning. Some found that minimum exposure increases familiarity and subsequent perceptions of accuracy, both immediately and after a week \cite{Pennycook}, whereas others found that such exposure does not materialise into a change in behaviour \cite{desaint}.

Another line of research explored the relation between the spread of misinformation in a particular geographic area and how that influences the behaviour or opinion of that population \cite{FORATI2021102473}. The challenge in such studies lies in accurately gauging the extent to which individuals were genuinely exposed to false information.
  

Tools for assessing an individual or a group's prolonged exposure to unreliable information on social media are very scarce and face considerable challenges due to limited data access. Regardless of the level and durability of the impact of misinformation on our behaviour, what is missing are technologies to quantify and monitor our long-term exposure to specific categories of misinformation, such as false claims about a particular political party or health issue.




\paragraph{Personalised Misinformation Corrections\\}

People, and the misinformation they consume, are not equal. We interact with different misinformation for different reasons and at different frequencies and intensities. To be effective with our online interventions, we should explore the potential of personalising our corrections to cater to these different experiences and personas. Technology can play a critical role in achieving this goal. 

Correcting individuals' beliefs in misinformation is a sensitive process that is influenced by numerous personal factors \cite{Ecker}. Technology can help detect various parameters at the individual level. For example, analysing the content someone shares on a social media platform can indicate their affiliation with known fake news sites, their previous exposure to false claims about specific topics, if and how they reacted to previous corrections, etc. Certainly, technologies capable of extracting such clues must undergo rigorous ethical scrutiny, even though the algorithms employed by the platforms to enhance user engagement often utilise very similar indicators.

In general, personalised misinformation correction is a promising yet complex approach, and much further research is needed to gain a comprehensive understanding of how technology can effectively implement and deliver it. 


\paragraph{Conclusions\\}


There is a big role that computing technologies can play in the fight against misinformation. Computer programs can be employed to assess whether a false claim or its corresponding fact-check is gaining more traction. They can also automatically assess and monitor the credibility of online accounts, and spot and even predict misinformation by leveraging rich semantic and statistical knowledge of previously debunked claims and their recurring templates. 

From historical spread patterns, machine learning can help determine more optimal times for releasing a fact-check for a particular misinformation. We can also measure long-term exposure to misinformation by analysing the timeline of online accounts. Bot-like programs can facilitate real-world scalable misinformation correction experiments, monitor audience reactions to corrections, and automatically tune and personalise interventions to maximise the visibility and impact of corrective information. 


The next frontier of combating misinformation lies in the responsible and ethical development of the technical capabilities above and firmly rooting them in the rich knowledge from psychology, sociology, and communication studies. This will drive the expansion of our research horizons and has the potential to yield more impactful solutions to the complex problem of misinformation.


\bibliographystyle{plainnat}

\end{multicols}
\end{document}